\documentclass[aps,prd,preprint,nofootinbib]{revtex4}
\usepackage{amsfonts}
\parskip + 5pt
\usepackage{epsfig}
\usepackage{graphicx}
\newcommand{\filetext}[1]{\texttt{#1}}
\newcommand{\shellcommand}[1]{\texttt{#1}}
\newenvironment{entry}%
{\begin{list}{}{\setlength{\topsep}{0mm} \setlength{\itemsep}{0mm}
\setlength{\parskip}{0mm} \setlength{\parsep}{0mm}
\setlength{\leftmargin}{20mm} \setlength{\rightmargin}{0mm}
\setlength{\labelwidth}{18mm} \setlength{\labelsep}{2mm}}}%
{\end{list}}

\newcommand{\iteme}[1]{\item[\texttt{#1}\hfill]}

\begin{document}
\title{DJpsiFDC:\\an event generator for the process $gg\rightarrow J/\psi J/\psi$  at LHC\\[7mm]}

\author{%
Cong-Feng Qiao$^{1,2}$%
\quad Jian Wang$^{1}$
 \quad Yang-heng Zheng$^{1}$ }
\affiliation{$^{1}$College of Physical Sciences, Graduate University
of Chinese Academy of Sciences \\ YuQuan Road 19A, Beijing 100049,
China} \affiliation{$^{2}$Theoretical Physics Center for Science
Facilities (TPCSF), CAS}

\author{~\vspace{0.7cm}}

\begin{abstract}
DJpsiFDC is an event generator package for the process
$gg\rightarrow J/\psi J/\psi$. It generates
 events for primary leading-order
$2\rightarrow2$ processes. The package could generate a LHE document
and this document could easily be embedded into detector simulation
software frameworks. The package is produced in Fortran codes.
\\

\noindent {\bf PACS numbers:} 02.70-c; 11.55.Hx

\noindent {\bf Keywords:} Event generator, FDC, $J/\psi$ pair,
color-octet scheme

\textbf{Program summary}

\emph{Program Title:} DJpsiFDC

\emph{Operating system:} Linux (with GNU FORTRAN 77 compiler)

\emph{Programming language:} FORTRAN 77

\emph{External libraries:} CERNLIB 2001 (or CERNLIB 2003)

\emph{Distribution format:} tar gzip file

\emph{Size of the compressed file:} 2.99M Bytes.

\emph{Classification:} 11.1

\end{abstract}

\maketitle

\section{Introduction}

NRQCD\cite{nrqcd} has now become a basic theory to parameterize
non-perturbative contributions represented by color-singlet(CS) and
color-octet(CO) matrix elements in heavy quarkonium production and
decays . The color-octet mechanism supplies a way to partly
understand the prompt $J/\psi$ and $\psi'$ surplus production at the
Fermilab Tevatron \cite{cdf,Braaten}. However, there are still
something unclear in this
scheme\cite{braaten02,b-factory-exclusive1,b-factory-exclusive2,
b-factory-inclusive1,b-factory-inclusive2,nloexclusive,nloinclusive,nlorelative}.
To investigate charmonium production mechanism has become an urgent
and important task in the study of quarkonium physics.

$J/\psi$ pair production process could be a candidate for testing
and verifying color-octet scheme. This process has an exclusive
multi-muon final state with extremely small backgrounds. But
according to the result of previous unpolarized
calculations\cite{barger,qiao1} for Tevatron experiment, the
cross-section of this process at Tevatron is very small, and
Tevatron experiment also has not supplied a solid evidence for
having detected it.

The running of the LHC provides a great opportunity to realize this
aim. With high center-of-mass energy of 14 TeV and luminosity of
about $10^{32}\sim10^{34}cm^{-2}s^{-1}$, results from the LHC
experiment may be able to answer how much color-octet mechanism
contributes to $J/\psi$ production under high collision energy. With
regard to the calculations of Qiao {\it et al} \cite{theory} and Li
{\it et al} \cite{pku}, the $J/\psi$ pair production process at LHC
could study charmonium production mechanisms. But firstly, we must
perform Monte Carlo simulation on this process to research its
experimental feasibility. Thus, we developed a generator package on
this process based on Feynman Diagram Calculation (FDC)
system\cite{fdc}. The generator contains only primary leading order
$gg\rightarrow J/\psi J/\psi$ channels, and we neglected those
channels such as $q\bar{q}\rightarrow J/\psi J/\psi$ which
contribute much less to than $gg\rightarrow J/\psi J/\psi$
processes. The output of the generator is a standard Les Houches
Event file\cite{lhe}, and could be embedded to the detector
simulation software framework.

The transverse momentum (Pt) distributions of muons from $J/\psi$
generated in color-singlet and color-octet schemes are shown here.
In large Pt region (Pt $>$ 10 GeV), the muons from $J/\psi$
generated in color-octet scheme are more than those from $J/\psi$
generated in color-singlet scheme. That means significant more
$J/\psi$ in color-octet scheme could be reconstructed in large Pt
region. If more $J/\psi$ events were detected in large Pt region,
then it could be considered that color-octet scheme contributes
significantly to the cross-section of the process. If the
cross-section was measured, the contribution of color-octet scheme
could also be determined.

\begin{figure}[b,m,u]
\centering
\includegraphics[width=10cm,height=8cm]{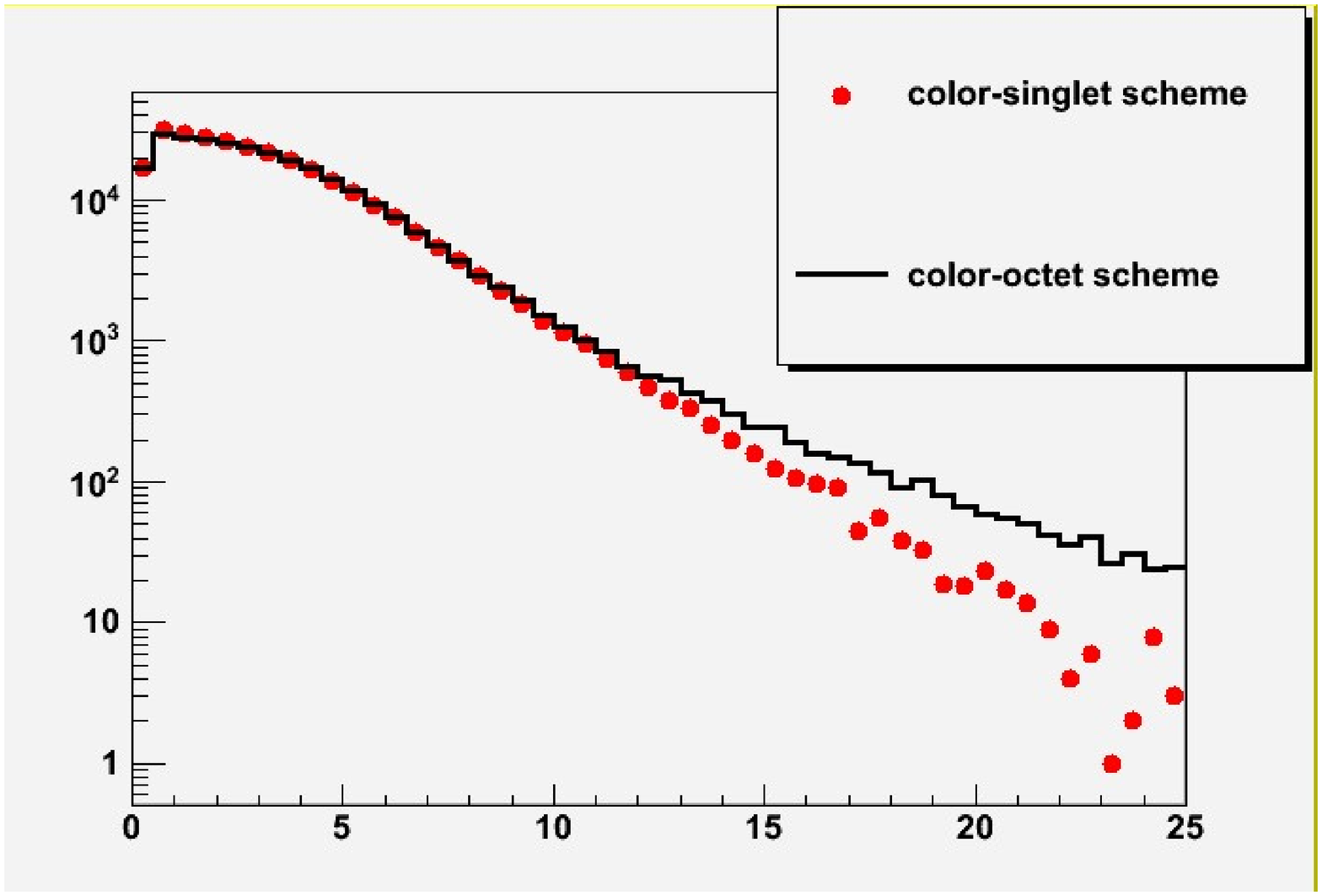}
\caption{\small Transverse momentum distribution of muon from
$J/\psi$ in color-singlet and color-octet schemes generated by
DJpsiFDC package.} \label{graph2}
\end{figure}

This paper is organized as follows: In Section 2, the directory
structures of the event generator package are described. In Section
3, we describe the details of the installation and implementation of
DJpsiFDC package. Sections 4 describes the format of the output
files. The summary is presented in Section 5.

\section{Directory
structures}\label{secSTRU}

There are five sub-directories in the directory. They are
\filetext{basesv5.1}, \filetext{DJpsi-generator}, \filetext{f77},
\filetext{gg2jpsi} and \filetext{octet$\_$gg2jpsi}. The directory
\filetext{basev5.1} stores \texttt{BASES} libraries used for
Monte-Carlo calculation of the cross-section integral. The directory
\filetext{f77} are the common-shared computational tools for physics
quantities, such as $\alpha_s$, etc, and the directory
\filetext{DJpsi-generator} contains the $J/\psi$ pair production
information.

\filetext{gg2jpsi} and \filetext{octet$\_$gg2jpsi} are two
directories containing color-singlet and color-octet parts of
$J/\psi$ pair production process respectively. All the compiled
executable programs and the input files are stored in these two
directories. In the following, the two parts would be described
sequentially.

\subsection{Color-Singlet Scheme}

%%%%%%%%%%%%%%%%%%%%%%%%%%%%%%%%%%%%%%%%Feynman-Diagram%%%%%%%%%%%%
\begin{figure}[b,m,u]
\centering
\includegraphics[width=10cm,height=8cm]{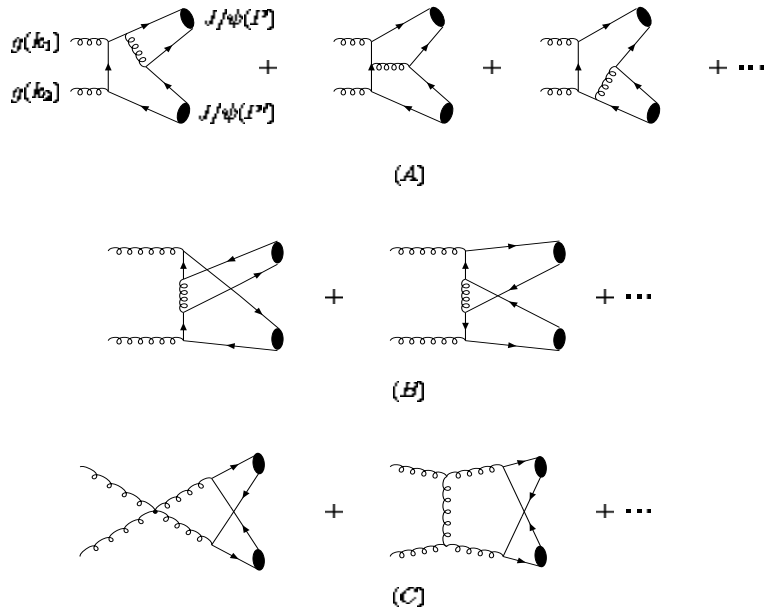}
\caption{\small Typical Feynman diagrams of $J/\psi$ pair production
in $p\,p$ collision at leading order in color-singlet scheme.}
\label{graph2}
\end{figure}

%%%%%%%%%%%%%%%%%%%%%%%%%%%%%%%%%%%%%%%%%%%%%%%%%%%%%%%%%%%%%%%%%%%

In color-singlet part,  we restrict simulation objects merely to the
gluon-gluon fusion one, as shown in Figure \ref{fig3}. In fact,
$J/\psi$ pair production process in color-singlet scheme includes $g
+ g \rightarrow J/\psi + J/\psi$ and $q + \bar{q} \rightarrow J/\psi
+ J/\psi$. Since $q + \bar{q} \rightarrow J/\psi + J/\psi$ processes
contribute much less, we did not consider this part's contribution.

\subsection{Color-Octet Scheme}

In the color-octet scheme, the final objects produced in hard parton
interaction are not $J/\psi$s, but color-octet bound states, as
shown in Figure \ref{fig4}. Here we comply with the convention in
PYTHIA system, considering this color-octet bound state as a new
particle, and assuming its invariant mass is 30MeV larger than
$J/\psi$, which is 3.126GeV. The color-octet bound state would
radiate a soft gluon, and decay to a real $J/\psi$.

It is worth noting that the $J/\psi$ pair hybrid production scheme,
i.e., one $J/\psi$ in a pair is produced though CS scheme but the
other is via CO scheme, is neglected in the calculation since its
contribution is comparatively very small.

%%%%%%%%%%%%%%%%%%%%%%%%%%%%%%%%%%%%%%%%Feynman-Diagram%%%%%%%%%%%%
\begin{figure}[b,m,u]
\centering
\includegraphics[width=10cm,height=4.5cm]{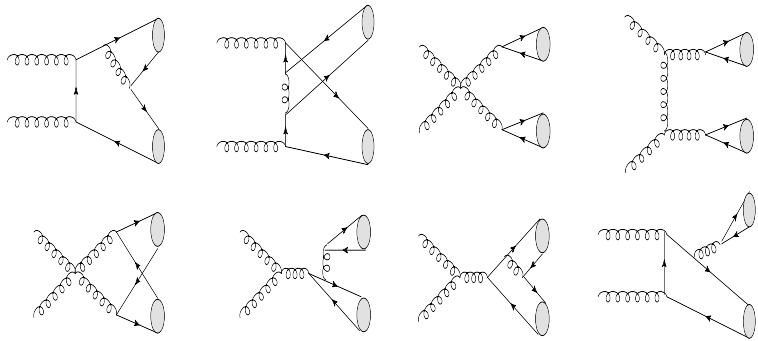}
\caption{\small Typical Feynman diagrams at leading order in
color-octet scheme.} \label{graph3}
\end{figure}

%%%%%%%%%%%%%%%%%%%%%%%%%%%%%%%%%%%%%%%%%%%%%%%%%%%%%%%%%%%%%%%%%%%

\section{Installation and Implementation}\label{secImp}

The generator is compressed in a file named \filetext{DJpsiFDC.taz}
and users could download it from:

\begin{quote}
\url{http://cid-4fc4170d6530982b.office.live.com/browse.aspx/.Public}
\end{quote}

Installation and implementation procedures are described
respectively below.

\subsection{Installation}

The installation steps are given below and for convenience we assume
that:
\begin{itemize}
\item[1.] the current user account is \texttt{user};
\item[2.] the current directory is \texttt{/home/user};
\item[3.] it is assumed that the user use \shellcommand{"csh"} ; if not , the user
should execute \shellcommand{"chsh useraccount" or "ypchsh
useraccount"} to change the shell class to cshell ;
\item[4.] User has installed CERN library and CERN library should be CERN2001.
\end{itemize}
For the installation, user first need to decompress the downloaded
gzip file by typing the following command.\\
   \hspace*{3em}\shellcommand{"tar -xzvf DJpsiFDC.taz"} \\
then home directory "DJpsiFDC" is generated.
Secondly, get into it by "cd DJpsiFDC" and set the correct path of CERN library in env.csh,then execute: \\
   \hspace*{3em}\shellcommand{ "source env.csh" and "installlib"}.  \\
It will do the installation.

Note: users must firstly modify the path and cern library location
in "env.csh" into their local paths, then execute "source env.csh"
so as to establish the environment smoothly.

\subsection{Implementation}

To implement the generator, user firstly needs to establish links to
program library. After entered home directory of generator, user
could execute \shellcommand{ "source env.csh" and "installlib"} to
realize it and this is a necessary procedure for the next
operations. Then user must enter color-singlet part or color-octet
part by executing "cd gg2jpsi/fort" or "cd octet$\_$gg2jpsi/fort".
Take color-singlet part as the example.

After changed current directory to gg2jpsi/fort, execute program
\filetext{make}, then three executable programs would appear. They
are:
\begin{table}[h]
 \centering
  \begin{tabular}{ll}
    \hline
    \filetext{int} & \hspace*{3em}calculates and stores the cross-section in file \filetext{fresult.dat}\\

    \filetext{gevent} & \hspace*{3em}generates and stores parton-level events information in file \filetext{pdata1.dat} \\

    \filetext{fdcpythia} & \hspace*{3em}generates and stores the final events information in file \filetext{DJpsiFDC.lhe} \\

    \hline
  \end{tabular}
\end{table}

Users only need to execute the executable programs to get generate
events and get output files, but user must call \filetext{int} ahead
to \filetext{fdcpythia}.

There are two input-files for users to change input parameters:
\filetext{input.dat} and \filetext{parameter.input}.
\filetext{input.dat} stores center-of-mass energy, and user could
generate events under different collision energy by changing it. In
\filetext{parameter.input}, user could configure initial parameters
such as event number to generate, etc. The format of
\filetext{parameter.input} is described as: each parameter occupies
one line; in each line, the parameter name and value are separated
by a equal sign (i.e. ``='') like:
\begin{verbatim}
    parameter = value;
\end{verbatim}
the line which begins with a number sign (i.e. ``\#") is recognized
as the comment. All parameters used by DJpsiFDC in
\filetext{parameter.input} are listed with brief explanation as:
\begin{entry}
\iteme{RUNNINGLEVEL(=1)} : To specify the running level.
It needs not to be cared here.

\iteme{EVNTNUM} : To specify the number of the generated events.

\iteme{PARTONEVNTNUM} : The number of the generated parton-level
events. It must be at least twice to EVNTNUM.

\iteme{PRESENTEVNTNUM} : The number of generated hadronic events in the last run.
It is set automatically and should not be modified by users.

\iteme{RESET} : To control the program \filetext{fdcpythia} to run in two
different modes -- \texttt{Re-generation} mode and
\texttt{Appending} mode. In DJpsiFDC package, we only use the
\texttt{Re-generation} mode (\texttt{=1}), since all the generated
events are erased, and new parton-level events and hadronic events
are generated.
\end{entry}

It should be noted that after changing collision energy in
\filetext{input.dat}, user must re-execute \filetext{int} to
calculate cross-section under new energy, or the generator would
generate events under original collision energy.

The generator could also change transverse momentum and rapidity
restrictions on the generated $J/\psi$ pair. Users could execute "vi
func.f", and at line 166-176 of func.f, they could find the
following codes:

 a = dsqrt(p(3,3)**2+p(2,3)**2)\\
 if(a.lt.4) out$\_$of$\_$physical$\_$region = .true.\\
 if(a.gt.40) out$\_$of$\_$physical$\_$region = .true.\\
 b = -dlog(tan(0.5*acos(dsqrt(p(3,3)**2\\
.    +p(2,3)**2+p(4,3)**2)**(-1)*p(4,3))))\\
 c = -dlog(tan(0.5*acos(dsqrt(p(3,4)**2\\
.    +p(2,4)**2+p(4,4)**2)**(-1)*p(4,4))))\\
 if(b.lt.-2.2) out$\_$of$\_$physical$\_$region = .true.\\
 if(b.gt.2.2) out$\_$of$\_$physical$\_$region = .true.\\
 if(c.lt.-2.2) out$\_$of$\_$physical$\_$region = .true.\\
 if(c.gt.2.2) out$\_$of$\_$physical$\_$region = .true.\\

a is the transverse momentum, and b, c are pseudorapidity of
respective $J/\psi$. Users could change the limitations by modifying
the numbers after \filetext{"$.lt.$"} or \filetext{"$.gt.$"}.
\filetext{"$.lt.$"} means \filetext{"less than"}, and
\filetext{"$.gt.$"} means \filetext{"greater than"}. After changed
limitations, users must re-execute \filetext{./make} to re-compile,
then operate as the above procedures.

\section{Output format}\label{secEVENTFILE}

There are two output files for users: fresult.dat and DJpsiFDC.lhe.
fresult.dat stores the calculated cross-section by the generator,
and user could use this value to estimate the number of events in
certain integrated luminosity.

DJpsiFDC.lhe is the primary output file storing main information of
original and final particles. "LHE" means Les Houches Event, its
content is identical with what was already defined by the Les
Houches Accord in 2003\cite{lhe}. This information is embedded in a
minimal XML-style structure. Complete file format looks like

$<$LesHouchesEvents version="1.0"$>$\\
 $<$!--\\
  $\#$ optional information in completely free format,\\
  $\#$ except for the reserved endtag (see next line) \\
  --$>$\\
  $<$header$>$ \\
  $<$!-- individually designed XML tags, in fancy XML style --$>$\\
  $<$/header$>$ \\
  $<$init$>$\\
  compulsory initialization information\\
  $\#$ optional initialization information \\
  $<$/init$>$ \\
  $<$event$>$\\
  compulsory event information \\
  $\#$ optional event information \\
  $<$/event$>$ \\
  (further $<$event$>$ ... $<$/event$>$ blocks, one for each event)\\
  $<$/LesHouchesEvents$>$

The LHE file has included all necessary information of original and
final parton-level particles, such as beam energy, color label,
particle ID, mom particle ID, four-momentum, invariant mass, etc.

\section{Summary}\label{secCON}

We have presented a FORTRAN program package DJpsiFDC which is a
 generator for simulating $J/\psi$ pair
production. It has implemented primary leading-order $2\rightarrow2$
processes and can generate events in color-singlet and color-octet
mechanism respectively. The generator would create a LHE document
containing initial and final particles information, which could be
embedded into detector software systems. This package has been
tested with setting the $pp$ collisions energy at $\sqrt{s}$=10TeV
and 14TeV, and the collision energy could be changed. Monte Carlo
simulation is being processed, and according to the actual LHC
operation status, the case under 7TeV is being studied. With data
accumulation on LHC, the conclusions on this process could be
attained in future two or three years.

%%%%%%%%%%%%%%%%%%%%%%%%%%%%%%%%%%%%%%%%%%%%%%%%%%%%%%%%%%%%%%%%%%%%%
\vspace{1cm}

{\bf Acknowledgments}

We are grateful to Guoming Chen and Jianguo Bian for numerious
stimulating discussions on the generator issue. This work was
supported in part by the National Natural Science Foundation of
China(NSFC) under the grants 10935012, 10928510, 10821063 and
10775179, by the CAS Key Projects KJCX2-yw-N29 and H92A0200S2, and
by 100 Talents Program of CAS.
%%%%%%%%%%%%%%%%%%%%%%%%%%%%%%%%%%%%%%%%%%%%%%%%%%%%%%%%%%%%%%%%%%%%%%

\end{document}